\documentclass[12pt]{article}
\usepackage{amsmath,amsthm,amssymb,euscript,epsf,epsfig}
\usepackage{array}

\newcommand{\be}{\begin{equation}}
\newcommand{\ee}{\end{equation}}
\newcommand{\bea}{\begin{eqnarray}\displaystyle}
\newcommand{\eea}{\end{eqnarray}}
\newcommand{\nn}{\nonumber}

\newcommand{\pd}{\partial}

\makeatletter
\@addtoreset{equation}{section}
\makeatother

\def\one{{\hbox{ 1\kern-.8mm l}}}
\def\zero{{\hbox{ 0\kern-1.5mm 0}}}

\def\co{ {\cal{O}} }  

\begin{document}

{}~
{}~
\hfill\vbox{\hbox{}\hbox{QMUL-PH-05-13}
}\break

\vskip .6cm

\centerline{{\Large \bf  Noncommutative geometry,  Quantum effects
                and  }}
\centerline{{ \bf \Large DBI-scaling  in the collapse
 of D0-D2 bound states.  }}

\medskip

\vspace*{4.0ex}

\centerline{\large \rm
C.~Papageorgakis ${}^1$, S.~Ramgoolam ${}^1$, N.~Toumbas  ${}^{2}$}

\vspace*{4.0ex}
\begin{center}
{\large ${}^1$Department of Physics\\
Queen Mary, University of London\\
Mile End Road\\
London E1 4NS UK\\
}
\vskip.1in 
{\large ${}^2$Department of Physics\\
University of Cyprus\\
Nicosia 1678, Cyprus \\}
\end{center}

\vspace*{5.0ex}

\centerline{\bf Abstract} \bigskip

We study fluctuations of time-dependent fuzzy two-sphere solutions 
of the non-abelian DBI action of D0-branes, describing 
a bound state of a spherical D2-brane with N D0-branes. 
The quadratic action for small fluctuations is shown to be identical
to that obtained from the dual abelian D2-brane DBI action, using 
the non-commutative geometry of the fuzzy two-sphere.    
For some of the fields, the linearized equations 
take the form of solvable Lam\'e equations. 
We define a  large-N DBI-scaling limit, with vanishing 
string coupling and string length, and where the  
gauge theory coupling remains finite. In this limit, the non-linearities 
of the DBI action survive in both the classical and the quantum context, 
while massive open string modes and closed strings decouple. 
We describe a critical radius where strong gauge coupling effects
become important. The size of the  bound  quantum ground state 
of multiple D0-branes makes an intriguing appearance 
as the radius of the fuzzy sphere, where the maximal angular momentum quanta become 
strongly coupled. 

\thispagestyle{empty}

\vfill
\begin{flushright}
{\it ${}^{\dagger}${\{c.papageorgakis , s.ramgoolam \}@qmul.ac.uk ,
 nick@ucy.ac.cy  }\\ }
\end{flushright}
\eject

\section{ Introduction } 

 We consider a time-dependent spherical $D2$-brane system
 with homogeneous magnetic flux.
 This is described by a fuzzy sphere solution 
 to the non-abelian action of 
 $N$ D0-branes or equivalently by an abelian $D2$-brane action. 
 The classical solutions 
 have been studied in the  context of Matrix Theory 
 and non-abelian DBI in \cite{kata,coltuck,rst, chenlu,papram,wath}.
 Related systems  involve $D1$$\perp$$D3$ brane intersections 
 \cite{gibbons,calmal,cmt}. 
 Equivalence at the level of classical solutions exists in a large 
 class of examples \cite{papram,cmm,cmti} including higher 
 dimensional fuzzy spheres \cite{clt,sphdiv,horam,Azbag,Kimura}. 
 It is natural to explore whether the equivalence at the level 
 of classical solutions extends to an equivalence at the level 
 of quadratic fluctuations. 

 In this paper we study the fluctuations of the 
 time-dependent $D0$-$D2$ brane system. 
 In section 2, we consider the 
 action for fluctuations using the $D2$-brane action. 
 We find that the result is neatly expressed 
 in terms of the open string variables of \cite{SW}. 
 The quadratic action is a ($U(1)$) Yang-Mills theory 
 with a time-dependent coupling, effective metric and a $\Theta$-parameter. 
 The radial scalar couples to the Yang-Mills gauge 
 field. We analyze the wave equation for the scalar fluctuations
 and identify a critical radius of the fuzzy sphere 
 where strong coupling effects set in. This radius
 is different for different values of the angular momentum of the excitations. 
 The fluctuation equation for scalars transverse 
 to the $\mathbb{R}^3$ containing the embedded sphere turns 
 out to belong to the class of solvable  Lam\'e equations. 
 It is very interesting that such an integrable 
 structure appears in a non-supersymmetric context. 

 In section 3, we obtain the quadratic action for fluctuations 
 on the sphere from the non-abelian symmetrised trace action \cite{tseyt} 
 of $N$ $D0$-branes.  
 We find precise agreement with the action obtained 
 from the $D2$-side. The fact that the commutators 
 $ [ \Phi_i , \Phi_j ] $ contain terms which scale differently 
 with $N$ means that we need to keep $1/N$ terms from commutators 
 of fields. The noncommutative geometry of the fuzzy sphere
 \cite{madore, iktw} is reviewed and applied to this derivation.   
 We observe that the mass term for the radial scalar 
 we obtain can also be calculated from the reduced action for the 
 radial variable. This simple calculation is extended to higher dimensional 
 fuzzy spheres and shows similar qualitative features.
 
 In section 4, we describe a DBI scaling
 limit, where $ N \rightarrow \infty$ , 
 $g_s \rightarrow 0$  and  $\ell_s \rightarrow 0 $ keeping fixed the 
 quantities $ L = \ell_s \sqrt{\pi N} $, $\tilde g_s = g_s \sqrt { N}$ 
 along with specified radius variables and gauge coupling constants.  
 In this limit, the non-linearities of the gauge coupling, 
 which have a square root structure coming from the 
 DBI action, survive. We discuss the physical meaning of this 
 scaling and its connection with the DKPS limit \cite{dkps},  
 which is important in the BFSS Matrix Model proposal for M-theory \cite{bfss}. 

 In section 5, we conclude with a discussion of some of the issues 
 and avenues related to the fluctuation analysis of the collapsing 
 $D0$-$D2$ system.

\section{ Yang-Mills type action for fluctuations }
When the spherical membrane is sufficiently large, we may use the 
Dirac-Born-Infeld (DBI) action to obtain a small fluctuations action 
about the time dependent solution of \cite{rst}. 
The DBI action is given by
\be
-{1 \over 4 \pi^2 g_s \ell_s^3} \int dt d\theta d\phi \sqrt{-\det
  (h_{\mu\nu} + \lambda F_{\mu\nu})}\;,
\ee
where $ \lambda = 2 \pi \ell_s^2 ~ $; $h_{\mu\nu}$ is the induced
 metric on the brane and 
$F_{\mu\nu}$ describes the gauge field strength on the membrane. 
The gauge field configuration on the brane consists of a 
uniform background magnetic field, 
$B_{\theta\phi}=N\sin \theta /2$, and the 
fluctuations $f_{\mu\nu}$: $F_{\mu\nu}=(B+f)_{\mu\nu}$. 
The background magnetic field results from the original $N$
$D0$-branes, which dissolve into uniform magnetic flux inside the $D2$-brane.

To quadratic order in the fluctuations, the action will involve 
a Maxwell field coupled together with a radial scalar field  
controlling the size and shape of the membrane. 
The parameters of this theory will be time-dependent because we are 
expanding about a time dependent solution to the equations of motion. 
For the radial field we write 
$\tilde{R}=R + \lambda(1-\dot{R}^2)^{1/2} \chi(t,\theta,\phi)$, 
where $R$ satisfies the classical equations of motion and 
$\chi$ describes the fluctuations. The normalization is chosen for 
later convenience. We also take into consideration scalar
fluctuations in the directions transverse to the  $\mathbb{R}^3$ containing the embedded 
$S^2$ of the  brane worldvolume, described by six scalar fields  
$\lambda\;\xi_m(t,\theta,\phi)$. 
Using the equations of motion, we 
have that the background field $R$ satisfies the following
conservation law equation \cite{rst}
\be\label{conservation}
1-{\dot R}^2 = {R^4 + N^2 \lambda^2/4 \over R_0^4 + N^2 \lambda^2/4} =
{R^4 + L^4  \over R_0^4 + L^4 }  \;.
\ee
We have introduced the physical length $L$ defined by 
\be 
L^2 = { N \lambda \over 2 } \;,
\ee 
which simplifies   formulas and plays an important role in the scaling 
 discussion of section 4. Here $R_0$ can be thought of as the initial radius of the brane at
which the collapsing rate $\dot R$ is zero.
The solution $ R(t)$
 to (\ref{conservation}) decreases from $R_0$ to zero,  
 goes negative and then oscillates back to its initial value. 
It was argued, using the $D0$-brane picture \cite{papram}, that the
 physical radius 
$R_{phys}$ should be interpreted as the modulus of $R$. 
Hence this is a periodic collapsing/expanding membrane, 
which reaches zero size and expands again. 
The finite time of collapse is given by
\be
\bar t =  c{\sqrt{R_0^4 + L^4 } \over R_0}\;,
\ee
where the numerical constant $c$ is given by $K(1/\sqrt{2})/\sqrt{2}$, with $K$
a complete elliptic integral.

To leading (zero) order  in the fluctuations, the induced metric
$h_{\mu\nu}$ on the brane is given by
\be
ds^2 = -(1-{\dot {R}}^2)dt^2 + R^2 (d\theta^2 + \sin^2\theta d\phi^2).
\ee
From the form of the induced metric we see that the proper time $T$
measured by a clock co-moving with the brane is related 
to the closed string frame time $t$ by a varying boost factor
\be
dt = {dT \over \sqrt{1 - {\dot R}^2}}.
\ee
So an observer co-moving with the collapsing brane concludes that the 
collapse is actually occurring faster.
In terms of proper time, the metric takes the form of a closed 
three-dimensional Robertson-Walker cosmology
\be
ds^2 = -dT^2 + R^2(T) (d\theta^2 + \sin^2\theta d\phi^2)
\ee
with scale factor $R$. The analogue of the Friedman equation is the conservation law  (\ref{conservation}).

Expanding the DBI action to quadratic order in the fluctuations we 
obtain the following:
\be\label{quadres}
S_{2}= -\int dt d\theta d\phi {\sqrt{-G} \over 2g_{YM}^2}\left[{1 \over 2}
G^{\mu\alpha}G^{\nu\beta}f_{\mu\nu}
f_{\alpha\beta} + G^{\mu\nu}\partial_{\mu}\chi\partial_{\nu}\chi + 
m^2 \chi^2 + G^{\mu\nu}\partial_{\mu}\xi_m\partial_{\nu}\xi_m\right]\;.
\ee
The effective metric $G_{\mu \nu }$ seen by the fluctuations is given by 
\be\label{openmet}
ds_{open}^2= -(1-{\dot R} ^2)dt^2 + {R^4 + L^4 
 \over R^2}d\Omega^2\;.
\ee
As we will see it is precisely the open string metric 
defined by \cite{SW} in the presence of background B-fields. 
The coupling constant is given by 
\be 
g_{YM}^2  = { g_s \over \ell_s} { \sqrt { R^4 + L^4 } \over R^2 },   
\ee
and the mass of the scalar field is given by
\be
m^2 = \frac{6 R^2}{(1-\dot R ^2 )\left(R^4+ L^4 \right)^2}
\left(  L^4 -R^4 \right)\;.
\ee
As expected, linear terms in the fluctuations
add to total derivatives once we use the equations of 
motion for the scale 
factor $R$. 

The set-up here differs   
from the original set-up of Seiberg/Witten 
\cite{SW} in that we have a non-constant B-field, $ B_{\theta \phi } =
  N \sin{\theta}/ 2  $. However the basic observation that in the presence of a background magnetic field, 
the open strings on the brane see a different metric $G_{\mu\nu}$  from the closed string frame 
metric $h_{\mu\nu}$ \footnote{More precisely, the metric $h_{\mu\nu}$ is induced on the brane
due to its embedding and motion in the background flat closed string geometry. Distances on the brane 
defined by using $h_{\mu\nu}$ are also those measured by
closed
string probes. Thus we shall call $h_{\mu\nu}$ the `closed string metric'.} 
\bea 
h_{00} &&= ~~ -(1-\dot R^2) \nn \\ 
h_{\theta \theta } &&= ~~  R^2 \nn \\ 
 h_{\phi \phi } &&= ~~ R^2 \sin^2{ \theta }
\eea 
continues to be true. The metric $G_{\mu\nu}$ is indeed related to $h_{\mu\nu}$ by 
\be
G_{00}= h_{00}, \, \, \,  G_{ab}=h_{ab}-\lambda^2 (Bh^{-1}B)_{ab}
\ee
or 
\be 
G_{\mu \nu } = h_{\mu \nu } - \lambda^2 (Bh^{-1}B)_{\mu \nu }.
\ee 
The open string metric (\ref{openmet}) is qualitatively
different from the closed string metric. 
Despite the fact that the original 
induced metric $h_{\mu\nu}$ becomes singular when the 
brane collapses to zero size, the open string metric $G_{\mu\nu}$ 
is never singular. To see this, let us compute the area 
of the spherical brane in the open string frame. This is given by
\be\label{area}
A = 4\pi \left( R^2 + {L^4 \over R^2}\right).
\ee
As $R$ varies, this function has a minimum at $R = L $, 
at which $A_{min}=4\pi N \lambda$ and the density of $D0$-branes
 is precisely at its maximum $1 / 4\pi \lambda$; that is, of 
order one in string units. Effectively, the open strings 
cannot resolve the constituent $D0$-branes at distance scales shorter 
than the string length.

The coupling constant can be expressed as $g_{YM}^2 = G_s
\ell_s^{-1}$, where
\be
G_s = g_s \left( {\det G_{\mu\nu} \over \det ({h_{\mu\nu} + 
\lambda B_{\mu\nu}})}\right)^{1/2}= g_s { \sqrt{R^4+ L^4  }\over R^2}.
\ee
So as $R$ decreases, the open strings on the brane eventually become 
strongly coupled. 

There is also a time-dependent vacuum energy density
\be\label{vacuum}
S_0= - {1 \over 4\pi^2 g_s \ell_s^3} \int dt d\theta d\phi \sqrt{-G} 
{R^2 \over \sqrt{R^4 + L^4 } }.
\ee
This vacuum energy density can be interpreted
 as the effective tension of the brane in the open string frame. 
In terms of the $D2$-brane tension $T_0=1/4\pi^2\ell_s^3$, this is given by
\be
T_{eff}= T_0 { R^2 \over \sqrt{R^4 + L^4 }}.
\ee 
We see that the brane becomes effectively tensionless as $R \to 0$.
  This is another indication that the theory eventually 
  becomes strongly coupled. The mass of the scalar
  field $\chi$ is a measure of the supersymmetry breaking scale of 
  the theory. Supersymmetry is broken because the brane is compact: the
mass tends to zero as $R \to \infty$.

There is a linear term 
\be
S_1={ 1 \over 2 \lambda^2}\int dt d\theta d\phi {\sqrt{-G} \over  
g_{YM}^2} \Theta^{ab}f_{ab}~~~  , \;
\ee
which is a total derivative, and can be dropped if 
we restrict to gauge fields of trivial first Chern class. 
It is noteworthy that the open string $ \Theta $  parameter, 
given by the standard formulas in terms of closed string frame parameters 
\cite{SW},  is precisely what appears here, 
\be
\Theta^{ab}= \lambda{\left( 1 \over h + \lambda B\right)}_A^{ab}.
\ee
In terms of $R$ this is given by
\be
\Theta^{\theta\phi}= - {  2 \over N } { L^4 \over (R^4 + L^4 )\sin\theta} .
\ee
The interpretation of $\Theta$ as a non-commutativity 
parameter will be made more clear in 
section 4. 
Notice that this attains its maximum value as $R \to 0$, at which
point $\Theta \sim 2/N\sin\theta$ being equal to the inverse
background magnetic field. 

In addition, there is a non-zero mixing term between the field strength $f_{\mu\nu}$
and the scalar field $\chi$ to quadratic order in the fluctuations. This is given by
\bea\label{mixing}
S_{int} &&=-\int dt d\theta d\phi {\sqrt{-G} \over \lambda g_{YM}^2}
 \frac{2 R^3}{\sqrt{1-\dot R^2} \left(\frac{\lambda^2 N^2}{4}+R^4\right)}\chi\Theta^{ab}f_{ab}  \nn\\
&&=  -\int dt d\theta d\phi {\sqrt{-G} \over  L^2 g_{YM}^2}
  \frac{ R^3}{\sqrt{1-\dot R^2} \left( L^4 +R^4\right)}\chi(N\Theta^{ab})
f_{ab}.\;
\eea
The second line makes it clear that this term is of 
order one if we consider the physical scaling limit\footnote{More on
  that in section \ref{qm}.} $\ell_s\rightarrow 0$, $N\rightarrow\infty$,
$g_s\rightarrow 0$ while keeping $R$ and $L$ fixed. Therefore, it is comparable 
to the other terms appearing in the fluctuation analysis.  
In performing various integrations by parts we have made extensive use of
the fact that the combination $ (\sqrt{-G} \Theta^{\theta\phi})/g_{YM}^2$ is given by
\be
 {\sqrt{-G} \over  g_{YM}^2} \Theta^{\theta\phi}=-\frac{\lambda^2 N 
   \ell_s}{2 g_s}\sqrt\frac{{1- \dot
     R^2}}{R^4 +\frac{\lambda^2 n^2}{4}}=-\frac{\lambda^2 N 
   \ell_s }{2 g_s} \frac{1}{\sqrt{R_0^4 + L^4   }}\;,
\ee
which is time-independent.

Thus, in the open string frame, the effective metric and 
non-commutativity parameter are well behaved all the way through the 
evolution of the brane. The coupling constant diverges as $R \to 0$. From the point 
of view of open string matter probes on the brane,
the sphere contracts to a finite size and then expands again as can be seen from eq. (\ref{area}). 
But the expansion results eventually in a strongly coupled phase.

The `open string' parameters $G_{\mu\nu}$, $G_s$ and $\Theta$ appearing in the 
above action are the ones which more naturally would appear in the 
description of the brane degrees of freedom in terms of
non-commutative field variables. We shall show in the next sections how such a description is realized
if we replace the smooth membrane configuration (and the uniform background magnetic field) 
with a system of $N$ $D0$-branes, and re-derive the
effective action for the fluctuations from the non-abelian DBI action of the
$D0$-brane system in the large-$N$ limit. In the $D0$-brane description the non-commutative variables are
$N\times N$ matrices; alternatively, the non-commutative variables can be expressed in terms of functions 
on a fuzzy sphere whose coordinates are non-commutative \cite{iktw}.  

One may turn off the scalar fluctuations $\chi$ and consider only 
fluctuations of the gauge field on spherical branes. 
In this set up one has a continuum fluid description of the $D0$-branes
on the collapsing brane. Indeed the gauge invariant 
field strength $F_{\mu\nu}$ describes the density and 
currents of the particles \footnote{Such fluid descriptions are given in the brane constructions of
\cite{Brodie,Freivogel}.}. This continuum description 
eventually breaks down for two reasons: Firstly the non-commutativity 
parameter increases, indicating that the fuzziness in 
area spreads over larger distances. Secondly the gauge field
fluctuations become strongly coupled.

\subsection{ Strong coupling radius }

Let us now determine the size of the brane at which the strong 
coupling phenomenon appears. 
First notice that the coupling constant $g_{YM}^2$ is dimensionful, 
with units of energy. Thus the dimensionless effective 
coupling constant is given by $g_{YM}^2/E_{proper}$, where
$E_{proper}$ is a typical proper energy scale of the 
fluctuating modes. The dependence of the effective coupling constant
on the energy reminds us that in $2+1$ dimensions the 
Yang Mills theory is weakly coupled in the ultraviolet and strongly 
coupled in the infrared. Because of the spherical 
symmetry of the background solution, angular momentum is conserved
including interactions. Thus as the brane collapses, we 
may determine the relevant proper energy scale in terms of the angular
momentum quantum numbers characterizing the 
fluctuating modes. 

To this end, let us examine the massless wave equation, as it arises
for example for the transverse scalar fluctuations
\be
\partial_{\mu}\left({\sqrt{-G} \over 
g_{YM}^2}G^{\mu\nu}\partial_{\nu}\xi\right)=0.
\ee
In terms of angular momentum quantum numbers, this becomes
\be\label{xi}
{1 \over (1-{\dot R}^2)}\partial_t^2\tilde\xi + {R^2l(l+1) \over 
(R^4 + L^4  )}\tilde\xi=0\;,
\ee
where we have set $\xi=\tilde\xi(t) Y_{lm}$ 
with $Y_{lm}$ being the appropriate spherical harmonic.

 The proper energy is given
 approximately by
\be
E_{proper}\sim {R\sqrt{l(l+1)} \over  \sqrt{(R^4 + L^4  )}}.
\ee 
As the brane collapses the wavelength of massless modes is actually 
red-shifted! This is essentially because of the form 
of the effective open string metric.

Now we let the brane collapse to a size $R \ll N^{1/2}\ell_s$. At smaller 
values of the radius the effective coupling 
constant becomes
\be\label{effective}
g_{eff}^2 \sim {g_s N^2 \ell_s^3 \over R^3 \sqrt{l(l+1)}}.
\ee
Clearly this becomes of order one when  $R$ approaches the strong coupling radius 
$R_s$ 
\be\label{rstrong}
R_s =  g_s^{1/3}\ell_s \left( {N^2 \over \sqrt{l(l+1)}}\right)^{1/3} 
= L  \left (  { g_s \sqrt {N} \over \sqrt { l(l+1) } } \right )^{1/3} \;.
\ee
Notice the appearance of $\ell_{11}=g_s^{1/3} \ell_s$, the  characteristic
scale of Matrix Theory. For $l$ close to the cutoff $N$,
 $R_s \sim N^{1 / 3} \ell_{11} $,  which is the estimated size of the 
quantum ground state of $N$ $D0$-branes \cite{pol, Susskind}. 
In general $R_s$ involves an effective  $N$ given  by
 $N_{eff} \sim N^2/\sqrt{l(l+1)}$.
We shall discuss these special values of the radius 
in more detail when we describe the membrane 
after taking various interesting limits
for the parameters appearing in (\ref{rstrong}).
 
The coupling constant of the theory (\ref{quadres})
 is time dependent. We can instead choose to work
with a fixed coupling constant absorbing the time-dependence solely in the 
effective metric if we perform a suitable conformal transformation. 
 By defining 
$\tilde{G}_{\mu\nu}=\Lambda G_{\mu\nu}$, the gauge field kinetic
 term gets multiplied by a
factor of  $\Lambda^{1/2}$.
Then we can 
re-define the coupling constant: $\tilde{g}_{YM}^2=g_{YM}^2/\sqrt{\Lambda}$. 
The conformal
transformation requires also suitable re-scalings of the
 fields $\chi$ and $\xi_m$ as well as appropriate
redefinitions of the various dimensionful parameters of the theory 
such as $m^2$
  and the non-commutativity parameter $\Theta^{ab}$.
\par
Choosing $\Lambda= (L^4+R^4)/ R^4$, the transformed coupling becomes
\be
{\tilde{g}_{YM}^2}=
{g_s\ell_s^{-1}},\;
\ee
and so it is time independent.
The open string metric in this frame is still non singular. However, 
the relevant dimensionless coupling is still the effective
coupling $g_{eff}^2$, eq. (\ref{effective}), which for small radii remains large. The effect of
the conformal transformation gets rid of the time dependence in the
coupling constant but also red-shifts $E_{proper}$ by a factor of $\Lambda^{-1/2}$. Therefore, we
cannot escape the strong coupling regime in this fashion.

\subsection{Overall transverse fluctuations and exactly solvable Schr\"odinger equation } 
Another interesting feature of (\ref{xi}) is that 
it is an integrable problem. 
Using $ ( 1 - \dot R^2 ) = ( R^4 + L^4 ) / (R_0^4 + L^4 ) $ 
the  wave equation becomes 
\be 
\partial_t^2 \tilde \xi + l(l+1){ R^2 \over  R_0^4 + L^4  }
 \tilde \xi  =  0 \;.
\ee 
Substituting the solution for the scale factor $R$, which is known in terms of 
the Jacobi elliptic function as 
$ R = R_0\; Cn \left( \frac{t\sqrt{2} R_0 }{\sqrt{R_0^4+L^4}}, { 1\over \sqrt {2} } 
 \right ) $, we have 
\be 
\partial_t^2 \tilde \xi + l(l+1) \frac{R_0^2}{R_0^4 + L^4} Cn^2 \left(\frac{t\sqrt{2} R_0 }
{\sqrt{R_0^4+L^4}} , { 1 \over \sqrt {2} }   \right ) \tilde \xi  = 0\;.
\ee

In \cite{papram} the solution to the classical problem is related  
to an underlying elliptic curve.
 For this specific case we can explicitly express
the Jacobi-$Cn$ function in terms of Weierstrass-$\wp$ functions of the
underlying curve\footnote{ The interested reader can find a 
  discussion of the Jacobi Inversion problem and its relevance to
  membrane collapse in \cite{papram} and references therein.
 The formulas that we present here
  can be checked by consulting Appendix C of that paper.
 A complete mathematical
  review of the Theory of Abelian functions can be found in \cite{enolrev}.}.
 The following
relation is true for this case
\be\label{minklame}
Cn^2\left(\sqrt{2}u,\frac{1}{\sqrt{2}}\right)=\frac{\wp(u;4,0)-1}{\wp(u;4,0)+1}\;.
\ee
For these specific functions the following identity
also holds
\be
\wp(u+\omega_3;4,0)=-\frac{\wp(u;4,0)-1}{\wp(u;4,0)+1}\;,
\ee
where $\omega_3$ is the purely imaginary half period of the relevant
elliptic curve in its Weierstrass form, and is given by
\be
\omega_3=i\int_0^1\frac{ds}{\sqrt{4s(1-s^2)}}\;.
\ee
After a re-scaling of time $t=u{\sqrt{L^4+R_0^4}}/{R_0}$ we
end up with
\be
\partial_u^2\tilde\xi +  l(l+1) Cn^2 \left( u \sqrt{2} , {1 \over \sqrt{2}} 
\right)  \tilde\xi 
= 
\partial_u^2\tilde\xi- l(l+1) \wp(u+\omega_3;4,0)\tilde\xi=0\;.
\ee
This is exactly the $g$-gap  Lam\'e equation for the ground state of
the corresponding one-dimensional quantum mechanical problem, which
has solutions in terms of ratios of Weierstrass $\sigma$-functions
( for an application in supersymmetric gauge theories 
 see for example  \cite{bakas} ). 

A related solvable Schr\"odinger problem arises in the one-loop computation of 
the Euclidean path integral. This requires the computation of 
the determinant of the operator
 \be
  - \partial_{\tau}^2  + { {R ( i \tau ) }^2 \over \sqrt { R_0^4 + L^4 } }
 l(l+1),   
\ee
 where we have performed an analytic continuation
$ t \rightarrow i \tau $ . The eigenvalues of the operator 
are determined by 
\be 
 - \partial_{\tau}^2 \tilde \xi  + { { R ( i \tau ) }^2 \over \sqrt { R_0^4 + L^4 } } 
 l(l+1) \tilde \xi = \lambda\; \tilde \xi \;.
\ee 
In \cite{papram} it is shown that $ R ( i \tau ) =  { 1 / R ( \tau ) }   
 $ and that $ R^2 ( i \tau ) = \wp ( \tau - \Omega ; 4 , 0 ) $ where 
$ \Omega = \int_0^1  { ds \over \sqrt{4 s ( 1 -s^2 ) } }.   $   Hence 
the eigenvalue equation becomes 
\be\label{euclame} 
- \partial_{\tau}^2 \tilde \xi   + l ( l+1) \wp ( \tau - \Omega ; 4 , 0 )
 \tilde \xi = \lambda\; \tilde \xi \;,
\ee 
where the eigenstates are also obtained in terms of $\sigma $-functions. 
\par
We postpone a detailed description and physical interpretation 
of the solutions of (\ref{minklame}) and (\ref{euclame}) for future work.  
It is intriguing that equation (\ref{minklame})
 has appeared in the literature on 
reheating at the end of inflation  \cite{inflation}. 
The physical meaning of this similarity, between fluctuation 
equations for collapsing D0-D2 systems and those of reheating, 
remains to be found.

\section{ Action for fluctuations from the zero-brane non-abelian DBI } 

The non-abelian DBI action 
for zero branes \cite{tseyt,myersdiel} is given by 
\bea\label{actzero}
 S = -{ 1 \over g_s \ell_s } \int dt \; STr \sqrt {  - \det ( M ) }   \;,
\eea 
where
\be 
M = 
\begin{pmatrix}
 & -1 & \lambda \partial_t \Phi_j    \\
                & - \lambda \partial_t  \Phi_i  & Q_{ij} 
\end{pmatrix}  
\ee 
and
\bea 
&& Q_{ij} = \delta_{ij} + i \lambda   \Phi_{ij}    \nn \\
&& \lambda = 2 \pi \ell_s^2  \;,
\eea 
with the abbreviation
\be
\Phi_{ij}=[\Phi_i,\Phi_j]\;.
\ee
The determinant of $M$, when the only non-zero  scalars lie
in the $i,j,k \in \{ 1,2,3 \} $ directions, is given by 
\bea\label{detexp}  
- \det{M}  &&= 1 + { \lambda^2 \over 2 } \Phi_{ij} \Phi_{ji} -  
\lambda^2 ( \partial_t \Phi_i )  ( \partial_t \Phi_i )   \nn \\
&& ~~~~ - { \lambda^4 \over 2 }  ( \partial_t \Phi_k ) ( \partial_t \Phi_k )
 \Phi_{ij} \Phi_{ji} + \lambda^4 (\partial_t \Phi_i)  \Phi_{ij} \Phi_{jk} 
 ( \partial_t \Phi_k ) \;.
\eea
These terms suffice for the calculation of the quadratic action for
the fluctuations involving the gauge field and the radial scalar. 
 However, when we include
 fluctuations  for the scalars  $ \Phi_m$ for $ m = 4 ... 9 $ 
we need the full $10\times 10$ determinant.
Fortunately, since we will
only be interested in contributions up to quadratic order, the relevant terms
will only be those of order up to  $ \lambda^4 $ 
\bea\label{trafl}
\nn &&{\lambda^2\over 2}\Phi_{im}\Phi_{mi}\left(1+
\frac{\lambda^2}{4}\Phi_{jk}\Phi_{kj} - \partial_t(\Phi_i )\partial_t (\Phi_i) \right)
- \lambda^2 \partial_t (\Phi_m) \partial_t(\Phi_m) 
\left( 1+\frac{\lambda^2}{2}\Phi_{ij}\Phi_{ji}\right)\\
 && -\frac{\lambda^4}{4}\Phi_{mi}\Phi_{ij}\Phi_{jk}\Phi_{km} +  \lambda^4
\partial_t(\Phi_i)\Phi_{im}
\Phi_{mj}\partial_t(\Phi_j)-\lambda^4 \partial^t(\Phi_m)
\Phi_{mi}\Phi_{ij}\partial_t (\Phi_j)\;  . 
\eea 
The expansion with terms of order up to $ \lambda^8 $
is given in \cite{cmm}.    

The $D2$-brane solution is described by setting 
$ \Phi_i = \hat R (t)  X_i $, where the matrices $X_i$
 generate the $N$-dimensional 
irreducible representation of $SU(2)$.
 By substituting this ans\"atz 
into the $D0$-action, we can derive equations of motion 
which coincide with those derived from the $D2$ DBI-action
\cite{rst}. In the correspondence we use
\be 
R^2 = \lambda^2 C ( \hat R )^2  \;,  
\ee 
where $C$ is the Casimir of the representation, $ C = N^2 $ in the 
large-$N$ limit. 
 Note that the square root form in the $D0$-action 
is necessary to recover the correct time of collapse. 
If we use the $D0$-brane Yang-Mills limit, we get the same functional 
form of the solution in terms of Jacobi-$Cn$ functions, but the time 
of collapse for initial conditions where $R_0$ is large is incorrect. 
The correct time of collapse increases as $R_0$ increases toward infinity,
whereas  the Yang-Mills limit gives a time which decreases in this limit. 
The need for the square root was realized in the context of spatial 
solutions $ \hat R ( \sigma )$ which describe $D1\perp D3$ funnels
\cite{cmt}.
We expand around the solution as follows
\bea\label{expn}
 \Phi_i &=& \hat R  X_i + A_i \nn \\
 A_i   &=& 2 \hat R  K_i^a A_a + x_i \phi \nn\\
 \Phi_m &=&  \xi_m \;.
\eea 
The decomposition in the second line above will be explained shortly. 
Throughout this section, we will be working in the $A_0=0$ gauge.

\subsection{ Geometry of fuzzy two-sphere : brief review } 
We review some facts about the fuzzy sphere 
and its application in Matrix theories\footnote{See for example 
\cite{madore,iktw}.}. 
As before, the $X_i$'s are generators of the $SU(2)$ algebra 
satisfying
\bea\label{algeb}  
[ X_i , X_j ]  = 2 i \epsilon_{ijk} X_k \;.
\eea 
With this normalization of the generators, the Casimir 
in the $N$ dimensional irreducible representation is given by $ X_i X_i = ( N^2 -1 )$.
If we define $ x_i =  { X_i/ N } $, we see that
\bea\label{coords}   
\nn  x_i x_i &=& 1  \\
  \lbrack x_i ,x_j \rbrack &=& 0
\eea
 in the large-$N$ limit. 
 Hence, in the large-$N$ limit the $x_i$'s reduce to Cartesian coordinates describing 
the embedding of a unit $2$-sphere in $ \mathbb{R}^3 $. For traceless symmetric  
tensors $ a_{j_1\ldots j_l } $ the functions 
$ a_{j_1\ldots j_l } x_{j_1}\ldots x_{j_l} $ describe spherical harmonics 
in Cartesian coordinates. Since general (traceless) Hermitian matrices can be expanded 
in terms of (traceless) symmetric polynomials of the $X_i$'s, 
hence in terms of the $x_i$'s\footnote {The latter 
give the correctly normalized spherical harmonics as we will explain later.}, 
all our fluctuations such as $A_i $ or transverse scalars  
such as $ \xi_m $ become fields on the sphere in the large-$N$ limit.
The expansion of  
$A_i$ is given by 
\be\label{expai} 
A_i = a_i + a_{i; j}x_j + a_{i; j_1 j_2} x_{j_1} x_{j_2} + \ldots \;.
\ee 
We can write this as $ A_{i} ( t , \theta , \phi ) $, with 
the time dependence appearing in the coefficients 
$a_i$, $a_{i; j_1\ldots  j_l }$ and the dependence on the angles arising from the polynomial 
of the $x_i$'s.  
 At finite-$N$, two important things happen: The $x_i$'s become non-commutative
and the spectrum of spherical harmonics is truncated at
$N-1$.  We will be concerned, in the first instance, with the large-$N$ 
limit.

The action of $X_i $ on the unit normalized coordinates
follows from the algebra (\ref{algeb}) 
\be 
[ X_i , x_j ] = 2 i \epsilon_{ijk} x_k 
\ee
and can be rewritten 
\be 
-2i \epsilon_{ipq}x_p \partial_q  ( x_j ) \;.
\ee 
So the adjoint action of $X_i $ can be written as
\bea\label{adact}  
&&[X_i , ~~ ] =   -2i K_i = -2i \epsilon_{ipq}x_p \partial_q \nn \\
&&          = -2iK_i^a \partial_a \;.
\eea
We have used  Killing vectors $K_i $ defined by 
$K_i = \epsilon_{ipq}x_p \partial_q $,  which 
 obey $ x_i K_i =0 $. They are tangential to the 
sphere and  can be expanded as $K_i^a \partial_a ~ $ , where 
$a$ runs over $ \theta , \phi $.  The components 
$K_i^a $ have been used in (\ref{expn}) to pick out 
the tangential gauge field components, and the radial component 
$ \phi  $ defined in (\ref{expn}) obeys 
$ \phi = x_i  A_i $.  
It is useful to write down the explicit components of $K_i$.  
The Killing vectors $K_i^a $ are given by 
\bea 
&&  K_1^{\theta} = - \sin \phi  \qquad   K_1^{\phi} = - \cot \theta \cos\phi 
   \nn \\ 
&&  K_2^{\theta} = \cos \phi  ~~ \qquad  K_2^{\phi} = -\cot \theta \sin \phi 
  \nn \\ 
&& K_3^{\theta}   =  0   ~~~~~~ \qquad  K_3^{\phi} =  1  \nn \;.
\eea 
Some useful formulas are the following 
\bea\label{useflf} 
 K_i^a K_i^b &=& \hat{h}^{ab } \nn\\ 
 x_i K_i^a   &=& 0 \nn\\ 
 K_i^a K_i^b \partial_a x_j \partial_b x_j &=&  2  \nn\\ 
 \epsilon_{ijk} x_i K_j^{a}K_k^{b} &=& {  \epsilon^{ab} \over \sin \theta }
= \omega^{ab} 
\eea
where $\epsilon^{\theta\phi}=1$. Here $\hat{h}_{ab} $ is the round metric
 on the unit sphere and $ \omega^{ab} $ is the inverse of the
 symplectic form.  
As a related remark, note that 
\be 
 \Theta^{ab } = -
{ \lambda^2 N \over 2 ( R^4 + { \lambda^2 N^2 \over 4} )}  
\epsilon_{ijk}x_i  K_j^a K_k^b  = - { 2\over N } { L^4 \over R^4 + L^4 } \epsilon_{ijk}x_i  K_j^a K_k^b  \;.
\ee

We will use these formulas to derive the action for the 
fluctuations $A_a $, $ \phi $ geometrically as a field 
theory on the sphere in the large-$N$ limit. 
We need one more ingredient. The $D0$-brane action is expressed 
in terms of traces, which obey the  $SU(2)$ invariance condition 
$ Tr ( \Phi ) = Tr [X_i , \Phi ]  $. This can be used to show that 
 if $\Phi $ is expressed as $ \Phi = a + a_j  x_j +
 a_{j_1 j_2} x_{j_1}  x_{j_2} + \cdots $, then the trace 
is just $ N a $, i.e. it picks out the coefficient of the trivial $SU(2)$
representation. By using the similar $SU(2)$ invariance property of 
the standard sphere integral we have 
\be\label{stdsph} 
 { Tr \over N }    ~~ \rightarrow ~~  { 1 \over 4 \pi } 
\int d \theta d \phi \sin \theta \;.
\ee
This relation between traces and integrals makes it clear why 
we have chosen the Cartesian spherical harmonics to 
be symmetric traceless combinations of $x_{i_1}\ldots  x_{i_l}
= ( X_{i_1} \ldots X_{i_l})/N^l  $. 
Such spherical harmonics obey $$ \int d \Omega  Y_{lm} 
Y_{l^{\prime} m^{\prime}} = 
{ Tr \over  N }  Y_{lm} Y_{l^{\prime} m^{\prime}}   = 
\delta_{l l^{\prime}}\delta_{m m^{\prime} } $$    
and are the appropriate functions to appear in (\ref{expai}).

 We make some further general  remarks on the calculation, before 
 stating the result for the action obtained from the 
 $D0$-brane  picture.  
 Note that the last term in the expansion of the determinant 
 (\ref{detexp}) 
 gives zero when we evaluate it on the ans\"atz $ \Phi_i = \hat R X_i $ 
 used to obtain the 
 solution, but it becomes non-trivial in calculating the action for the 
 fluctuations $ \Phi_i = \hat R X_i + A_i $. 
 The zero appears because the symmetrised trace allows us to  
 reshuffle the $X_i$ with the $[ X_i,X_j]$ for example. 
 Using this property and the commutation relations gives the desired 
 zero and hence leads to agreement between the effective actions 
 for the radial variable, as derived from the $D2$-brane picture.

\subsection{ The action for the gauge field and radial scalar } 
 
Using the ans\"atz (\ref{expn}) we have
\be\label{ijcom}   
[ \Phi_i , \Phi_j ]  =  
 ( \hat R )^2 [ X_i , X_j ] + \hat R  [X_i ,A_j ] + \hat R [A_i , X_j ] 
 + [ A_i, A_j ]     \;.
\ee
The first term scales like $N$, the second two terms 
are of order one in the large-$N$ limit, while the 
last term is of order $ {1/ N } $. 
The last commutator term is sub-leading in $1/N$ since the $ x_i$'s appearing 
in (\ref{expai}) commute in the strict large-$N$ limit, as of (\ref{coords}).
When computing terms such as the potential term 
$\sim [\Phi_i , \Phi_j]^2$, it is important to note that 
there are terms of order one coming from squaring 
$ \hat R  [X_i ,A_j ] + \hat R [A_i , X_j ] $ as well as 
from the cross terms $ ( \hat R )^2 [ X_i , X_j ][ A_i, A_j ] $. 
 For this reason, the underlying non-commutative geometry 
 of the fuzzy $2$-sphere is important in deriving even the 
 leading terms in the dynamics of the fluctuations.

The first term  in (\ref{ijcom})
is simplified by using the commutation relations 
to give $ 2i ( \hat R )^2 \epsilon_{ijk} X_k = 
2iN ( \hat R )^2 \epsilon_{ijk} x_k $. The second term 
can be written as $ -4i  ( \hat R )^2 K_i^a \partial_a ( K_j^b A_b ) 
-2 i K_i^a \partial_a ( x_j \phi )  $ 
using (\ref{adact}).
We can compute
the leading $1/N$ correction arising from the commutator
 $[A_i, A_j]$ as follows. We can think of the
unit normalized, non-commuting coordinates $x_i$ as 
quantum angular momentum variables.
Since their commutator is given by $[x_i, x_j]=(2i\epsilon_{ijk}x_k)/N$,
 the analogue 
of $\hbar$ is given by $2/N$, which scales like the 
 inverse of the spin of
the $SU(2)$ representation. Thus the large-$N$ limit is equivalent to the
classical limit in this analogy, and in this case all matrix commutators $[A,B]$ can be approximated with
`classical' Poisson brackets as follows
\be\label{comm}
[A, B] \to {2i \over N}\left\{A,B\right\},
\ee
where $\{A,B\}$ is the Poisson bracket defined by
\be
\{A,B\}=\omega^{ab} \partial_a A\;\partial_{b}B\;,
\ee
using the inverse-symplectic form
appearing  in (\ref{useflf}).
As a check note that $\{x_i, x_j\}=\epsilon_{ijk}x_k$. The commutator $[A_i, A_j]$ is then given by
\be\label{noncomm}
[A_i, A_j]={2i \over N}\left\{A_i,A_j\right\}+O(1/N^2)={2i \over N}\omega^{ab}\partial_{a}A_i
\partial_{b}A_j + O(1/N^2).
\ee

Substituting in (\ref{actzero}) and expanding the square root, keeping up to
quadratic terms in the field strength components 
$F_{ab} = \partial_a A_b - \partial_b A_a $ and $F_{0a} = \partial_t A_a $, we obtain  
\bea 
-\int dt d\theta d\phi {\sqrt{-G} \over 4g_{YM}^2} 
G^{\mu\alpha}G^{\nu\beta}F_{\mu\nu}
F_{\alpha\beta}\;,
\eea 
where the effective metric and coupling constant are the ones 
appearing in section 2. Hence we have recovered from the D0-brane action
(\ref{actzero}) the  first term 
of (\ref{quadres}) obtained from the 
small fluctuations expansion of the $D2$-brane DBI action. 
We remark  that in calculating the quadratic term 
in the spatial components 
of the field strength, the last term in (\ref{detexp}) 
gives zero, but its contribution is important in getting 
the correct coefficient in front of $F_{0a}^2$. 

There is a term linear in $F_{ab}$ given by
\be
S_1={ 1 \over 2 \lambda^2}\int dt d\theta d\phi {\sqrt{-G} \over  
g_{YM}^2}r^4 \Theta^{ab} F_{ab}
\ee
where $r$ is the dimensionless radius variable, $r =  R/L$. 
This differs from the linear term obtained from the $D2$ DBI
action by the $r^4$ factor, but the whole term is a total derivative. 
As such it vanishes in the sector where the fluctuations 
 do not change the net monopole charge of the 
background magnetic field. This is a reasonable restriction to 
put when analyzing small fluctuations around a monopole configuration. 

At first sight we could also have
 $A_a^2$ contributions, which would amount to a mass for
the gauge field. Such terms coming from $\pd_t\Phi_i
[\Phi_i,\Phi_j][\Phi_j,\Phi_k]\pd_t\Phi_k$ cancel among
themselves. The contributions from the other three terms of (\ref{detexp})
cancel each other up to total
derivatives, upon expanding the square root and
also performing partial integrations in both the spatial and time
directions. Here, we need to use the equation of motion for the scale factor $R$ or $\hat{R}$.
Some useful formulas are given in the appendix A. 
It is important to note that this mass term only vanishes 
if we keep the terms $ [X_i,X_j][A_i,A_j]$, which are order one 
terms obtained by multiplying the order $N$  with the  order $1/N$
 commutators. 
\par
Next we turn to fluctuations involving the scalar field $\phi$. 
The spatial part of the relativistic kinetic term is 
\be 
-{ \ell_s \over 2 g_s } \int dtd\theta d \phi \sin  \theta 
 { (2\lambda N \hat{R}^2) \hat{h}^{ab} \partial_a \phi \partial_{b} \phi 
\over \sqrt { (1 - \lambda^2N^2\hat{\dot R}^2)( 1 + 4\lambda^2N^2\hat{R}^4) } } 
 \;.
\ee
This agrees with the $D2$-calculation (\ref{quadres})
 if we make the natural identification 
$ \phi = ( 1 - \lambda^2N^2\hat{\dot R}^2 )^{1/2} \chi $. 
Following this, we can match the 
quadratic terms in $\partial_t \phi $ and we find again that we get the same 
answer as from the $D2$-side. The overall kinetic term is given by
\be\label{kinetchi} 
- \int dtd\theta d\phi {\sqrt{-G} \over 2g_{YM}^2} 
 G^{\mu\nu} \partial_{\mu} \chi \partial_{\nu} \chi \;
\ee
as in (\ref{quadres}). 

For the mass term of $\chi$, we get 
\bea 
 &&-{N \over 4\pi\ell_s g_s} \int dt d \theta d \phi \sin{\theta}
\frac{12 \lambda^2 \hat R^2 (1-4 N^2\lambda^2 \hat R^4)}{(1+4
  N^2\lambda^2 \hat R^4)^{3/2}\sqrt{1-N^2\lambda^2 \hat
    {\dot R}^2}}\chi^2 \nn \\
&=& -\int dt d\theta d\phi   \frac{\sqrt { - G }}{2g_{YM}^2}
{6  R^2  (L^4 -  R^4)\over (L^4 +  R^4)^2 (1- \dot{ R}^2) } \chi^2  \;.
\eea
This agrees with the mass for $\chi$ in (\ref{quadres}).
 Another thing to note here is
that the determinant will also give contributions linear in $\phi$ and
$\pd_t \phi$ and also terms quadratic in the scalar fluctuation of the
form $\phi\;\pd_t \phi$. However, upon the expansion of the square root
to quadratic order
the overall linear factor of $\phi$ cancels. We recall that we are expecting the latter, since
$\phi$ is a fluctuation around a background which 
solves the equations of motion.  Upon conversion to
the $\chi$ variable the kinetic term for $\phi$ will also contribute
$\chi \; \pd_t \chi$ terms. Then
by integrating by parts and dropping the respective total time
derivative terms we end up with the appropriate mass for $\chi$ given above.

For the mixing terms between $F_{ab}$ and
$\phi$,  collecting all the relevant terms one gets
\be
-\int dt d\theta d\phi \frac{\sqrt{-G}}{g_{YM}^2}\frac{8 \hat R^3 N}{(1+4 \lambda^2 N^2 \hat
  R^4)\sqrt{1-\lambda^2 N^2 \dot{\hat R}^2}}\chi \Theta^{ab}F_{ab} \;.
\ee
Once more, we get exact agreement with the $D2$-calculation (\ref{mixing}).
Finally the quadratic action for the scalars $ \xi_m$
obtained by expanding the
terms in (\ref{trafl})  is easily   seen to be
\be
-\int dt d\theta d\phi \frac{\sqrt{-G}}{g_{YM}^2}G^{ab} \partial_a \xi_m
\partial_b \xi_m\;,
\ee
which  agrees with (\ref{quadres}). 

\subsection{Scalar fluctuations for the reduced action}
We expect to be able to reach the same results for the scalar
fluctuations by just considering the large-$N$ reduced action for the background
fields as in \cite{rst,papram}
\be\label{redact}
S_2=-\frac{2}{g_s\ell_s\lambda}\int dt \sqrt{1-\dot R^2} \sqrt{R^4
  +\frac{N^2 \lambda^2}{4}},
\ee
and consider adding fluctuations $R\rightarrow R+\lambda \sqrt{1-\dot R^2}\;\chi$ as before.
One gets 
\bea\label{massterm}
 S_2^\textrm{mass} && =-\frac{2}{g_s\ell_s\lambda}\int dt \frac{\lambda^2
  R^2 (3L^4-3R^4)}{2(L^4+R^4)^{3/2}\sqrt{1-\dot R^2}}\chi^2 \nn \\ 
&& =  { -4 L \sqrt { \pi } \over \tilde g_s  }  
         \int dt \frac{ 
  R^2 (3L^4-3R^4)}{2(L^4+R^4)^{3/2}\sqrt{1-\dot R^2}}\chi^2 
\eea
the same answer for the mass of the scalar
fluctuation as by perturbing the full action (\ref{actzero}),  when
written in terms of $g_{YM}^2$ and $\sqrt{-G}$.

We can make use of this result to check the behavior of the scalar
mass for higher even spheres. 
The reduced action for the fuzzy
$S^4$ is \cite{papram}
\be
S_4=-\frac{4}{g_s\ell_s \lambda^2 N}\int dt \sqrt{1-\dot
  R^2}\left(R^4+\frac{\lambda^2 N^2}{4} \right)\;.
\ee
Perturbing this will result to a mass
\bea
S_4^\textrm{mass} & =&-\frac{4}{g_s\ell_s \lambda^2 N}\int dt
\frac{2\lambda^2 R^2(3L^4-5R^4)}{(L^4+R^4)\sqrt{1-\dot R^2}}\chi^2 \nn \\ 
&=& -{  8 \sqrt { \pi } \over  { \tilde g_s } L } \int dt   
\frac{ R^2(3L^4-5R^4)}{(L^4+R^4)\sqrt{1-\dot R^2}}\chi^2,
\eea 
where we have made use of the appropriate equations of motion.

There is a similar behavior for the $S^6$. The reduced action is \cite{papram}
\be
S_6=-\frac{8}{g_s\ell_s\lambda^3N^2}\int dt \sqrt{1-\dot
  R^2}\left(R^4+\frac{\lambda^2 N^2}{4} \right)^{3/2},
\ee
and the result for the mass
\bea
 S_6^\textrm{mass}&=&-\frac{8}{g_s\ell_s\lambda^3N^2}\int dt
\frac{12\lambda^2 R^2(3L^4-7R^4)}{\sqrt{L^4+R^4}\sqrt{1-\dot R^2}}\chi^2\nn \\
&=& - {  48 \sqrt { \pi } \over \tilde g_s L } \int dt  
\frac{ R^2(3L^4-7R^4)}{\sqrt{L^4+R^4}\sqrt{1-\dot R^2}}\chi^2. 
\eea

The physical behavior remains the same for any $k$: for the pure
$N=0$ case the scalar mass squared is
negative from the beginning of the collapse all the way down to
zero. At finite (large) $N$ there is a transition point which
depends on the dimensionality $k$.

\subsection{ $1/N$ correction to the action } 

The derivation of the the action from the $D0$-brane side can easily 
be extended to include $1/N$ corrections. The net outcome will be a 
non-commutative gauge theory, where products 
are replaced by suitable star products. Two important features have to 
be noted. It is no longer consistent to assume 
$ x_i K_i^a = K_i^a x_i $. This is because 
\bea 
&& [ x_i , K_i^{\theta}] =  -{ 2i \over N } \cot{ \theta }  \nn \\ 
&& [ x_i , K_i^{\phi} ] = 0 \;.
\eea 
We can instead only assume $ x_i K_i + K_i x_i = 0 $. 
We also have a first correction to the Leibniz rule for the 
partial derivatives 
\be
 \partial_{a} ( FG ) = ( \partial_a F )G + F ( \partial_b G ) - 
 { i \over N } ( \partial_a \omega^{bc} )  ( \partial_b F ) ( \partial_c G ) \;.
\ee 
This is consistent with 
\bea 
\partial_a [ x_i , x_j ] = [ \partial_a x_i , x_j ] + [ x_i , \partial_a x_j ] 
                   - {  i \over N }  ( \partial_a \omega^{bc} )  [ \partial_b x_i , \partial_c x_j ] \;.
\eea

\section{ Scaling limits and Quantum Observables }\label{qm}

 Given the action we have derived from the 
$D0$ and $D2$-sides, there are several limits to consider 
so as to describe the 
 physics. 

\subsection{ The DBI-scaling } 

Consider $g_s \rightarrow 0 $ , $\ell_s \rightarrow 0 $ , $ N \rightarrow \infty $ 
keeping fixed 
\be\label{scaling1}  
 R ,\, \,  L = \ell_s\sqrt {  \pi N }   , \, \, g_s\sqrt { N} \equiv \tilde g_s\;.
\ee
In this limit the following parameters appearing in the 
Lagrangian are fixed 
\bea 
&&  g_{YM}^2 = { g_s \over \ell_s } 
{  \sqrt { R^4 + L^4 }  \over R^2  }  =  
  { \sqrt { \pi} \tilde g_s \over L }{  \sqrt { R^4 + L^4 }  \over R^2  } \nn \\ 
&&  G_{00} = \sqrt { 1 - \dot R^2 } \nn \\
&&  G_{ab} =  { { R^4 + L^4 } \over R^2 }  \hat{h}_{ab} \;.
\eea 
We also keep fixed in this limit the energies 
and angular momenta of field quanta in the theory. 

With this scaling all the quadratic terms of the field theory 
action on $S^2$ derived from the D2-brane side in (\ref{quadres}), (\ref{mixing}), 
and reproduced in section 3 from the D0-branes, remain fixed.  
 Notice that all terms in (\ref{detexp})
are also of order one and all of them contribute so as to obtain the small
fluctuations action and the parameters of the theory given above. In addition, 
since in this limit $\ell_s \rightarrow 0 $,  massive 
open string modes on the branes decouple, and we can neglect higher derivative corrections 
to the DBI action. Further,  since  $g_s \rightarrow 0 $, we expect  
closed string emission to be negligible. 
This scaling should be compared to scalings studied in
Matrix Theory in \cite{dkps,pol,sen,seimat,bgl,hks,imsy}. In the region  
 $R \ll L $, we will consider the relation to the Matrix Theory limit below. 

There are several interesting features 
of the limit (\ref{scaling1}). It allows us to neglect the 
finite size effects of the quantum $D0$-brane bound state. 
The quantum field theory we have derived by expanding around 
the classical solution might be expected to be invalid 
in the regime where the radius of the sphere reaches 
the size  $ R_{q}$  of the quantum ground state of $N$ $D0$-branes.  
This has been estimated to be \cite{pol, Susskind} 
\bea 
R_{q} &=& N^{1 / 3 } g_s^{1/3} \ell_s  \nn \\ 
      &=& { \tilde g_s L \over N^{1/3}} \;.
\eea 
Clearly this is zero in the scaling limit, which gives us reason 
to believe that the DBI action is valid all the way to $R=0$. 

Another issue is gravitational back-reaction. 
This can be discussed by comparing the radius 
of the collapsing object to the gravitational 
radius of a black hole with the same net charge. This type of  argument 
is used for example in \cite{klst} for studying collapsing domain walls 
in four dimensions. 
We find that in the scaling limit (\ref{scaling1}), gravitational back-reaction is negligible. 
To see this consider first the excess energy $ \Delta E $ 
of the classical configuration above the ground state energy of $N$ $D0$-branes. 
For extremal black holes the horizon area is zero. 
For non-extremal ones, it is directly determined 
by the excess energy \cite{klebtseyt}
\be 
R_h^8 = g_s^{25 \over 14} \ell_s^{121 \over 14 } \sqrt { N }
 ( \Delta E )^{9 \over 14 }. 
\ee 
Using 
\bea 
 \Delta E &=& { N \over g_s \ell_s   } \left (  
{ \sqrt { R_0^4 + L^4 } \over L^2 } -1  \right )  \nn \\
&=& { N^2 \over { \tilde g_s} L } \left (  
{ \sqrt { R_0^4 + L^4 } \over L^2 } -1  \right ) ,
\eea 
we find for the horizon radius 
\be 
R_h^8  = N^{ -24 \over 7 } { \tilde g_s }^{8 \over 7 } L^8 
\left (  { \sqrt { R_0^4 + L^4 } \over L^2 } -1  \right )^{9 \over 14} \;,
 \ee 
which goes to zero in the large-$N$ limit. 
This shows that gravitational back-reaction resulting in 
the formation of non-extremal black holes does not constrain
the range of validity of the DBI action in our scaling limit. 

Another black hole radius we may compare to 
is the Schwarzschild radius for an object having 
energy $ {N \sqrt{ R_0^4 + L^4 }/ (g_s \ell_s L^2) } $, as is the case
for our membrane configuration. 
This comparison is more relevant in the limit $ R \gg L$ where 
the $D0$-brane density is small; in other words the charge density of the relevant black hole 
is small.
In this case we expect 
the discussion of \cite{klst} to be most relevant. 
The Schwarzschild radius  is given by $ R_{sch} = ( G_s E )^{1/7 } $, 
or more explicitly  
\be 
R_{sch} = N^{-3 \over 7 } L { \tilde g_s }^{ 1 \over 7 } 
 \left ( \sqrt { R_0^4 + L^4 } \over L^2 \right )^{1\over 7}   ~~ . 
\ee 
This is also zero in the scaling limit (\ref{scaling1}), 
and hence does not invalidate the DBI action. 

Since $R$ is time-dependent, the parameters of the theory are also time-dependent. 
We may consider  correlators of gauge invariant operators 
\be 
\langle \co ( t_1 , \sigma^a_1 ) \co ( t_2 , \sigma^a_2 ) \ldots \rangle
\ee  
where $\co$ can be for example $ Tr ( F^2 )$ or $Tr ( \Phi^2 ) $,
which use the field strength or transverse scalars.  
For  times $t_1, t_2 , \ldots $ corresponding 
via the classical solution to $R$ near  $R_0$, the Yang-Mills coupling
is small, and the approximation where non-linearities of the DBI have been 
neglected is a valid one. So we can compute such correlators perturbatively. When $R$ approaches zero, the 
Yang-Mills coupling diverges, so we need to use the all-orders 
expansion of the DBI action. We have not computed the fluctuation action to all 
orders, but it is in principle contained in the full DBI action.

An interesting observable is $ \langle 0 | \chi |0 \rangle $ which gives 
quantum corrections to the classical path. 
In time dependent backgrounds, one can typically define 
distinct early and late
times vacua because positive and negative frequency modes at early and late 
times can be different.
If we set 
up an early times vacuum in the ordinary manner, and write $\chi$ as a 
linear combination of early times creation and annihilation operators,
the one point function of $\chi$ in the late times vacuum may be non-zero
indicating particle production. The non-trivial relation between 
in and out-vacua is certainly to be expected for all the fields 
in the theory, since it is a generic feature of quantum fields 
in a time dependent background \cite{birdav}. Recent applications 
in the decay of unstable branes include \cite{strom,llm,uvfin}.

We have argued that radiation into closed string states is negligible because
their coupling constant $g_s \to 0$ in the scaling limit (\ref{scaling1}).
In the context of open string tachyon condensation, describing brane decay, 
the zero coupling limit 
of closed string emission was shown not to approach zero as 
naively expected because of a divergence coming from 
a sum over stringy states \cite{llm}. Here we may hope to 
escape this difficulty because $\ell_s \rightarrow 0 $ means that 
the infinite series of massive closed string states decouple
and the Hagedorn temperature goes to infinity. Of course in the tachyonic 
context \cite{llm}, the limit $\ell_s\to 0$ could not be taken since it would 
force the tachyon to be infinitely massive as well. 
To prove that there is no closed string production 
will require computation of the one-loop partition function 
in the theory expanded around the solution and showing 
 that any non-vanishing imaginary part obtained in 
 the limit (\ref{scaling1})  can be interpreted in terms 
of the DBI action (\ref{actzero}). Such computations 
in a supersymmetric context are familiar in Matrix Theory. 
Recent work has also explored the non-supersymmetric context
\cite{mw}.

We have argued that open strings on the membrane eventually become 
strongly coupled
when the physical radius is given by eq. (\ref{rstrong}). This
 special value for the radius remains fixed in 
our scaling limit: $R_{s} \sim (\tilde g_s)^{1/3}L$. It can
 be made arbitrarily small if we take
$\tilde g_s$ sufficiently small. But for any fixed value of this
 coupling, however small it is, 
strong coupling quantum effects are eventually needed to understand
the subsequent membrane evolution. Quantum processes may cause the
 original brane 
with $N$ units of $D0$-brane charge to
split into configurations of smaller charge. However such
 non-perturbative phenomena
should be describable within the full non-abelian $D0$-brane
 action (\ref{actzero}).

We can also construct multi-membrane configurations.
For example, we can construct $m$ coincident spherical membranes 
if we start with the non-abelian DBI action of $mN$ 
$D0$-branes and
replace the background values of the
matrices $\Phi_i$ in (\ref{expn}) with the following block-diagonal 
forms \cite{iktw}
\be
\hat{R} X_i \to \hat{R} X_i \otimes \one_m.
\ee  
The fluctuation matrices $A_i$ are replaced by
\be
A_i \to  \sum_{1}^{m^2} A_i^{\alpha} \otimes T^{\alpha}
\ee
where the $m\times m$ matrices $T^{\alpha}$ are generators of $U(m)$. 
Taking the large-$N$ limit, while keeping 
$m$ fixed, the action for the fluctuations should result in 
a non-abelian $U(m)$ gauge theory on a sphere describing
a collection of $m$ coincident spherical $D2$-branes. 
The field strength of the $U(1)$ part of this gauge group attains a 
background value corresponding
to $mN$ units of flux on the sphere.  
We expect the effective metric and coupling constant of this
theory to be given by the same formulas that we have derived before. 
Separate stacks of $D2$-branes
can be constructed by giving an appropriate vacuum expectation value to
 one of the 
transverse scalars; that is, by `Higgsing' the $U(M)$ gauge group.
 The net background magnetic flux 
should now split appropriately among the separate stacks.  
Within this set-up, one can study non-perturbative instanton processes
 that result into transferring of 
$D0$-branes from one membrane stack to another, as in \cite{polpou}.
 The effective 
dimensionless coupling of such processes is given approximately by
 $g_{YM}^2/ \langle\phi\rangle$, where
$\langle\phi\rangle$ is the relevant Higgs VEV. When the branes are large,
 that is $R > L$, this coupling can be
kept small if we take $\tilde g_s$ small, and such processes are
 exponentially suppressed. 
But when the radius becomes small, the theory becomes
strongly coupled and such non-perturbative processes become relevant.

\subsection{ The D0 Yang Mills (Matrix Theory) limit } 

In this limit, we take $R/L=r$ as well as $r_0$ to be small. 
We will show how  
the effective action for the fluctuations in this regime can be
 derived from the BFSS Matrix Model \cite{bfss}. Earlier 
work on this model appears in \cite{halpern,dwhn}.

The effective parameters of the theory are $G_{\mu\nu}$,
 $G_s$ and $\Theta^{ab}$. 
In terms of the dimensionless radius
variable $r$, these are given by 
\bea
&&G_{00}=  - \sqrt{{r^4+1 \over r_0^4 +1}}, \ \ \   
G_{ab}= {N\lambda \over 2}\left(r^2 + {1 \over r^2}\right)\hat{h}_{ab} \nn \\
&& G_s=g_s {\sqrt{r^4 + 1} \over r^2} \nn \\
&& \Theta^{ab}=-{2\epsilon^{ab} \over N(1+r^4)\sin\theta} \;.
\eea
When $r, r_0 \ll 1$, these take the following `zero slope' form \cite{SW, Seiberg}  
\bea\label{zslope}
&&G_{00} \to \tilde{G}_{00}=1, \ \ \   
G_{ab}\to \tilde{G}_{ab}=-\lambda^2 (Bh^{-1}B)_{ab}= {N\lambda \over 2r^2}\hat{h}_{ab} \nn \\
&& G_s \to \tilde{G}_s=g_s\det(\lambda B h^{-1})^{1 \over 2}={g_s \over r^2} \nn \\
&& \Theta^{ab} \to \tilde{\Theta}^{ab}=(B^{-1})^{ab}=-{2\epsilon^{ab} \over N\sin\theta} 
\;.
\eea
Notice that the rate of collapse $\dot R$ is given by
\be
\dot R^2 = r_0^4 - r^4 
\ee
in this regime. In particular, this remains small throughout the collapse of the brane.

We can `derive' these zero slope parameters from the effective action of the constituent D0-branes 
in the small-$r$ regime. The background fields scale as 
\bea
&& \Phi_i = \hat{R}X_i =\left( {r \over 2L } \right)X_i \nn \\  
&& \partial_t \Phi_i = (\partial_t \hat{R})X_i \sim \left({\sqrt{r_0^4-r^4} \over 2L^2}\right)X_i. \;
\eea
We assume a similar scaling behavior for the fluctuations $A_i=2\hat{R}K_i^aA_a+x_i\phi$ and their
time derivatives $\partial_tA_i$ in the small-$r$ regime. That is, 
we take the gauge field $A_a$ to be of order one while
the radial fluctuations $x_i\phi$ to be at most of the order $r/L$ in magnitude. Similarly, the velocity
fields $\partial_tA_a$ and $x_i\partial_t\phi$ are required to be of order $r/L$ and $r^2/L^2$ respectively.
This is a reasonable requirement for the behavior of the fluctuations so as to keep them smaller or at least
comparable to the background values of the fields.   
Then the full fields $\Phi_i$ and their time derivatives are sufficiently 
small in the small-$r$ regime, and the $D0$-brane effective action (\ref{actzero}) 
takes the form of a 
$0+1$ dimensional Yang-Mills action:
\be\label{D0YM} 
S= {(2 \pi)^2 \ell_s^3 \over g_s} \int dt \left[Tr\left( {1 \over 2}\partial_t \Phi_i \partial_t \Phi_i + 
{1 \over 4}[\Phi_i, \Phi_j]^2 \right)
- {N \over \lambda^2 }\right].
\ee 
The second and third terms in (\ref{detexp}) scale as $r^4$ in the limit, while
the last two terms as higher powers of $r$. In the small-$r$ regime, we can neglect the last two terms of
(\ref{detexp}) and expand the square root of the DBI action dropping higher powers of $r$.
We end up with an action that is quadratic in the time derivatives of the fields and quartic in the fields
themselves \footnote{A similar expansion can be consistently carried out for the fields $\Phi_m$ that are transverse
to the $\mathbb{R}^3$ where the membrane is embedded.}. 
Roughly speaking, in this regime each $D0$-brane is moving slowly enough so that the non-relativistic, small velocity 
expansion of the DBI Lagrangian can be applied ending up with (\ref{D0YM}). 
This expansion is valid if we choose the initial radius parameter $r_0$ to be small enough, or
the initial physical radius to satisfy $R_0 \ll L$. Essentially the Yang Mills regime is valid when the
effective separation of neighboring 
D0-branes is smaller than the string scale throughout the collapse of the brane. 
Finally, in this regime the equation of motion for the scale 
factor is given by 
\bea
\ddot{\hat R}+8\hat{R}^3 &=& 0 \nn \\
 \ddot{r}+{2 \over L^2}r^3 &=& 0. \;
\eea

Setting $\Phi_i = \hat{R}X_i + A_i$, we can determine 
a matrix model for the fluctuating fields $A_i$. This matrix model
is equivalent to a non-commutative $U(1)$ Yang Mills theory on a fuzzy sphere \cite{iktw}. This correspondence maps 
hermitian matrices
to functions on the sphere, and replaces the matrix product with a suitable non-commutative star product.
 To see how the non-commutative gauge fields arise, we examine the transformation of the fluctuating matrices
$A_i$ under time independent infinitesimal $U(N)$ gauge transformations,
which are symmetries of the action (\ref{D0YM}).
Under such a gauge transformation, the matrices $\Phi_i$ and $A_i$ transform as follows
\bea
&& \delta_{\lambda}\Phi_i=i[\lambda, \Phi_i] \nn \\
&& \delta_{\lambda}A_i = -i\hat{R}[X_i, \lambda]+i[\lambda, A_i], \;
\eea
with $\lambda$ an $N \times N$ Hermitian matrix.
Using equation (\ref{adact}), the corresponding function on the sphere transforms as\footnote{We do not 
use different notation to distinguish the $N \times N$ hermitian matrices from their corresponding 
functions on the sphere. We hope the distinction is made clear from the context.} 
\be\label{nctrans}
\delta_{\lambda}A_i=2\hat{R}K_i^{a}\star\partial_{a}\lambda +i(\lambda \star A_i - A_i \star \lambda),
\ee
where $\lambda$ is now taken to be a local function on the sphere. Thus we end up with a $U(1)$ non-commutative gauge
transformation.
The gauge covariant field strength is given by
\be\label{ncf}
F_{ij}=i\hat{R}[X_i, A_j]-i\hat{R}[X_j, A_i]+i[A_i, A_j]+2\hat{R}\epsilon_{ijk}A_k=i[\Phi_i, \Phi_j]+
2\hat{R}\epsilon_{ijk}\Phi_k.
\ee
The last equation makes gauge covariance manifest. The field strength $F_{ij}$ is 
zero when the fluctuations are set to zero, while the commutator $[\Phi_i, \Phi_j]$ 
attains a background expectation value given by 
$\hat{R}^2[X_i, X_j]$. 

In the commutative limit, the non-commutative gauge transformations (\ref{nctrans}) reduce to ordinary 
local $U(1)$ gauge transformations. As we already discussed, 
this is equivalent to a large-$N$ limit. 
Decomposing 
$A_i=2\hat{R}K_i^{a}A_{a}+ x_i\phi$  
we see that in the commutative limit, 
the tangential fields $A_{a}$ transform
as the components of a gauge field on the sphere, $\delta_{\lambda}A_{a}=\partial_{a}\lambda$, while the
transverse field $\phi$ as a scalar. The full non-commutative gauge transformation  (\ref{nctrans}) 
though mixes $\phi$ and 
the vector field $A_{a}$ \cite{iktw}; this is another manifestation of the fuzziness of the underlying space.

It is easy to see that in the commutative limit the field strength reduces to
\be
F_{ij}\to 4\hat{R}^2 K_i^{a}K_j^{b}F_{ab}+ 2\hat{R}(x_jK_i^{a}-x_iK_j^{a})
\partial_{a}\phi -2\hat{R}\epsilon_{ijk}x_k\phi.
\ee
The deformation arising from the underlying non-commutativity comes from the commutator piece $i[A_i, A_j]$ 
in (\ref{ncf}). Up to the order of $1/N$, this deformation is given by (\ref{noncomm}), and can be rewritten as
\be
i[A_i, A_j]=\tilde{\Theta}^{ab}\partial_{a}A_i
\partial_{b}A_j + O(\tilde{\Theta}^2).
\ee 
We conclude immediately that the underlying non-commutativity parameter is $\tilde{\Theta}$. 

We can expand the $D0$-brane Yang Mills action (\ref{D0YM}) to quadratic order in the fluctuations in the
large-$N$ limit. Having
established the equivalence of the full $D0$-DBI action with the $D2$-brane action to this order in the
fluctuations, all we need to do is to replace the effective metric, coupling constant and non-commutativity
parameter with their `zero slope' values (\ref{zslope}). Of course, one can carry out the expansion directly 
using the action (\ref{D0YM}) and verify that the parameters of the theory in this regime are indeed 
given by $\tilde{G}_{\mu\nu}$,
$\tilde{g}_{YM}^2$ and $\tilde{\Theta}$.
The mass of the scalar field $\chi$ defined above eq. (\ref{kinetchi}) is given by 
\be
m^2 = {6r^2 \over L^2}
\ee
in this regime and it is positive.
Finally, the mixing term becomes
\be
-\int dt d\theta d\phi {\sqrt{-\tilde{G}} \over \tilde{g}_{YM}^2}{r^3 \over L^3} 
\chi (N\tilde{\Theta}^{ab})F_{ab}.
\ee 

It is important to realize that non-linearities in the equations of motion, arising from interaction terms of 
higher than quadratic order in the non-relativistic Lagrangian (\ref{D0YM}), 
are all suppressed by factors of $1/N$. From
the point of view of the $U(1)$ non-commutative field theory on the fuzzy sphere, all interaction terms
arise from the non-commutative deformation of the field strength (\ref{ncf}) and they end up being proportional
to powers of $\tilde{\Theta}$. It is easy then to see that non-linearities become important at
angular momenta of order $l \sim N^{1/2}$ where $ \tilde \Theta^{ab}
\partial_a \otimes \partial_b $ 
is of order one.   This fact was also emphasized in the
analysis of \cite{kabat}. From (\ref{rstrong}) then we see that such angular momentum modes become strongly
coupled when
\be\label{rstrong1}
R \sim \ell_{11}N^{1/ 2}
\ee 
or
\be
r \sim {g_s}^{1/ 3}.
\ee
Roughly, the strong coupling phenomenon occurs when in the closed string frame each $D0$-brane occupies an area
of order $\ell_{11}^2$, smaller than $\ell_s^2$.

In the scaling limit (\ref{scaling1}), the eleven dimensional Planck length tends to zero like $N^{-2/3}$ and 
the strong coupling radius (\ref{rstrong1}) goes to zero. 
Thus in the limit 
(\ref{scaling1})
the evolution of such small branes, described by the $D0$-brane Yang Mills action (\ref{D0YM}), 
can be treated classically throughout the collapse of the brane. 
We can alternatively
take a different scaling limit so as to probe the short eleven dimensional Planck scale, which sets
the distance scale at which strong coupling quantum phenomena occur in our system in the non-relativistic regime. 

We can take $g_s \to 0$ keeping $R$ and $\ell_{11}$ fixed, and also $N$ fixed and large. 
In this limit $L \to \infty$ like $g_s^{-1/3}$, so that $r$ and also $r_0$ are small. 
The physical field variables
$\lambda \Phi_i \sim  { R X_i/ N} $, and so they remain fixed in this limit. 
The same is true for their conjugate momenta. At the same time each
individual $D0$-brane is getting very heavy since $m_{D0}=1/g_s\ell_s \sim g_s^{-2/3}/\ell_{11}$.
Hence the $D0$-branes are slowly moving in this limit. 
This limit is the famous DKPS limit \cite{dkps,pol}  
in which the short distance scale probed by the $D0$-branes is the eleven dimensional Planck scale. 
Closed strings decouple
from the brane system. The same is true for excited massive open strings on the branes. 
This is because the energy of the
fluctuating massless open string states is much smaller than the mass of excited open string oscillators
in the limit \cite{dkps,pol} and so massive open strings cannot get excited. 
Finally, in the BFSS limit \cite{bfss} where the eleven circle radius is decompactified, the membrane
we constructed is just a boosted spherical M-theory membrane. 

The strong coupling phenomenon above
occurs at a physical radius which is bigger than the size of the bound 
quantum ground state (of the $N$ $D0$-branes) by a factor of $N^{1/6}$. However
angular momentum modes of order the cutoff $N$ become strongly coupled when
the physical radius $R$ becomes comparable
to the size of the ground state $\ell_{11}N^{1/3}$ as can be seen from (\ref{rstrong}).
It is interesting that this scale which is expected to emerge from a
 complicated ground state solution of the $D0$-brane Yang-Mills 
Hamiltonian also appears in the analysis of the linearized fluctuations 
of fuzzy spheres. 

There is yet another simple way to see the $\ell_{11} $ length 
scale. It involves the application of the Heisenberg 
uncertainty principle to the reduced radial dynamics. 
 The momentum conjugate to $R$ coming from the reduced action
 (\ref{redact}) is 
\be
\Pi_R=\frac{\sqrt{R_0^4-R^4}}{g_s \ell_s^3 \pi }\;.
\ee
 With $ (\Delta R)  \; \Pi_R>\hbar ~ $ and 
$\hbar\sim 1 ~ $,  we get
\be
 ( \Delta R ) > \frac{g_s \ell_s^3 \pi}{\sqrt{R_0^4-R^4}} \;.
\ee
Evaluating the uncertainty at $R=0$ and assuming 
 the whole trajectory
  lies within the quantum regime, i.e. $ R_0 \sim \Delta R $, we obtain 
 a critical value for the initial radius $R_0 \sim R_c$ 
 where 
 $R_c\sim g_s^{1/3} \ell_s$, which is the eleven dimensional Planck scale. 
This simple analysis does not detect the $N^{1/3} $ factor that appears 
in the more complete analysis above.

The above discussion has focused on the region where $ R $
 is much smaller than $L$. 
The region of $ R \gg L $ or equivalently $ L = \ell_s \sqrt {\pi N}  \rightarrow 0 $ is 
also of interest. In the strict $ N = 0 $ limit we have a $D2$-brane 
without $D0$-brane charge. The negative sign of the mass of the field $\chi$ that appears in (\ref{massterm})
for $ R >  L $ also appears in the problem of fluctuations around the 
pure $D2$-brane solution. 
This negative sign indicates that the zero mode of the field $\chi$ is tachyonic in this regime.
When $R_0$ is larger than $L$, the tachyonic mass 
naively causes an exponential growth for the zero mode of the fluctuation $\chi$.
At this point, higher order corrections to the action
involving the zero mode would become significant. However, 
the reduced action for the scalar dynamics has no exponentially growing solutions.
This means that higher order terms stop this exponential growth. In fact,
as $R$ crosses $L$, the sign of the mass changes and we go into an oscillatory 
phase. This transition is reminiscent of a similar transition which occurs 
in the equation for fluctuations in inflationary scenarios, see for example 
\cite{branmart}.  
In the case $ R_0 \le L  $ the time evolution of the radial fluctuation does not encounter 
the tachyonic region.

\subsection{Mixing with graviton scattering states } 

The key observable in the BFSS Matrix theory limit 
is the scattering matrix of $D0$-brane bound ground states 
made of $N_1, N_2, \ldots N_i$ $D0$-branes, where $N_i$ are all large. 
Since these interactions are governed by $\ell_{11} $, which goes 
to zero in the scaling limit (\ref{scaling1}), such interactions among such states
become irrelevant. However a simple estimate suggests that 
these states can mix with the fuzzy sphere states. 
Consider an $SU(2) $ representation of spin $j$ with 
$ N = 2j+1$. Consider also matrices  
\be 
U_{\pm}  = \begin{pmatrix}   
& 0 & 0 \nn \\
& 0 & b \pm iv 
\end{pmatrix}  \;.
\ee
The diagonal blocks 
are of size $ N_1 \times N_1 $ and $ ( N-N_1) \times (N-N_1)$. 
There are also the standard $N \times N $ $SU(2)$ matrices 
$J_+ , J_- $,  which act in this representation. In the fuzzy sphere 
configuration we set $ X_{\pm} \equiv ( X_1 \pm i X_2 ) =  J_{ \pm} $ while
in the scattering configuration we set $ X_{\pm}  = U_{\pm} $. 
We calculate $ Tr ( [J_{+} , U_{-} ] [J_-,U_+] ) $ and find 
this proportional to $ \lambda^2 ( \hat R )^2 N ( N-N_1) $.
 If $N_1$ is a finite fraction of $N$ then this goes like 
$ \lambda^2 ( \hat R )^2 N^2 \sim ( \hat R )^2 L^4  $
 in the large-$N$ limit, which is of the same order as the terms in the quadratic 
action for the fluctuations we have computed. This indicates 
that the collapsing membrane can undergo transitions to these 
scattering states and conversely the scattering states 
can give rise to membranes. 
 
\section{ Discussion and Outlook }

The discussion of the scaling limit in section 4 is 
very reminiscent of similar scaling limits in the context of 
BFSS Matrix Theory \cite{bfss} and the AdS/CFT duality \cite{malda}.
 The difference 
is that here we are keeping the non-linearities 
coming from the  non-abelian $D0$-brane action (\ref{actzero}). 
This action is of course less understood 
than the $0+1$ SYM of the BFSS Matrix Model or the $3+1$ SYM of the 
canonical AdS/CFT correspondence. 
For example a completely satisfactory supersymmetric 
version has yet to be written down, although 
some progress on this has been discussed in \cite{tseyt}. 
However it is significant that our scaling discussion of section 4.1 
highlights the fact that the appropriate supersymmetrised non-abelian 
DBI  action should provide a complete quantum mechanical description 
of the collapsing  $D0$-$D2$ system.

This may appear somewhat surprising, but we will 
argue not  unreasonable. 
When $R$ is close to $R_0$ we have a Yang-Mills 
action at weak coupling, and quantum correlation 
functions can be computed in a weak coupling expansion. 
In the strict large-$N$ limit,  the Yang-Mills theory is commutative, while 
$1/N$ corrections amount to turning the background 
sphere into a non-commutative sphere. 
When the correlation functions are localized in regions 
where the radius is small, the Yang-Mills coupling is large 
and non-linearities in the fields become important. 
There must be a quantum mechanical framework which 
provides the continuation of the correlators to this region. 
Since $\ell_s$ has been taken to zero, massive string modes 
decouple and the only degrees of freedom left to quantize are those 
that appear in (\ref{actzero}). String theory loops 
degenerate to loops of the fields in this action. 
The  conjecture suggested by these arguments is that
the fate of the collapsing $D0-D2$  system in the zero radius 
region, and in the regime of parameters of section 4, is contained in the 
quantum version of the supersymmetrised non-abelian DBI 
of (\ref{actzero}). We have outlined a framework for 
calculating quantum corrections to the classical bouncing path, 
and discussed processes where one membrane splits into multiple membranes, 
or membranes mix with scattering states made of large bound states 
of zero branes.

It will be interesting to look for a gravitational 
dual for the decoupled gauge theory of section 4. One possibility 
is to start with a time dependent multi-$D0$ brane solution
of the type considered in \cite{gibbfreed,shirash}. 
Then a $D2$-brane could be introduced as a perturbation, 
as the $D5$-brane was introduced in a background of $D3$ branes 
in \cite{polstrass}. The gravitational dual 
may shed light on the strong coupling regime 
of $ R \rightarrow 0 $.  Another approach for a spacetime gravitational 
description is to consider the spherical $D0$-$D2$ system as a 
spherical shell which causes a discontinuity in the 
extrinsic curvature due to its stress tensor, and acts as 
a monopole source for the two-form field strength due to the $D0$-branes, 
and a dipole source for the four-form field strength due to the 
spherical $D2$-brane. As long as the  $D0$-brane fluid description is 
valid, it should be possible to view the $D0$ branes as smeared on a 
sphere of time-dependent radius. Exploring the solutions and 
 regimes of validity of these different gravitational descriptions
 will undoubtedly be a very interesting avenue for the future. 
The recent paper \cite{csv} is an example where a gauge theory dual 
in a time-dependent set up is proposed.

Any discussion of (\ref{actzero}) will certainly remind 
many readers of symmetrised trace issues, such as those 
raised by \cite{hashtaylor}. The system considered here 
belongs to a class of configurations which 
come in families labeled by a size $N$ of matrices 
where $[  \Phi_i , \Phi_j ] $ goes to zero in the large-$N $
limit. Higher even dimensional fuzzy spheres and co-adjoint 
orbits also belong to this class. In these cases 
the large-$N$ limit often has some sort of abelian 
geometrical description. For the leading large-$N$ in these 
cases the ordering of Matrices 
does not matter and, at the level of classical equations 
of motion, the system can be compared with an abelian dual 
\cite{cmm,cmti}. Here we have extended the comparison to 
fluctuations and found agreement. It will be interesting to 
see if the comparison can be extended to higher orders 
in $1/N$, where an appropriate star product 
is used on the higher brane and the Matrix product on the $D0$-brane 
side is interpreted as a star product on the sphere. 
A successful identification will require the correct implementation 
of the symmetrised trace at higher orders. Given the subtleties 
of separating field strengths and derivatives in the non-abelian 
case ( discussed for example in \cite{bilal} ),
it is probably best to approach the question of symmetrised 
trace and its corrections by embedding the non-abelian system of interest 
into a family of systems labeled by a matrix size $N$ which 
which can be taken to be large and admits an abelian limit.

In this paper we have focused on the analysis of fluctuations 
in the case of time dependent $D0$-$D2$ solutions. A similar analysis can be 
performed in the spatial $D1$ $\perp$ $D3$ configurations. 
Some aspects of this problem have
already been studied in \cite{bhatmell}. In the case of systems 
involving higher dimensional fuzzy spheres, such as $D0$-$D4$
 ($D1$$\perp$$D5$) systems or 
$D0$-$D6$ ($D1$$\perp$$D7$) systems,
 we expect on general grounds that there will 
be an abelian description based on a geometry of the form 
$ SO(2k+1)/U(k)$ and a non-abelian description on the 
$S^{2k}$ \cite{sphdiv,horam,nick1, nick2}. A detailed  fluctuation
 analysis of the kind studied here should allow a more precise description 
 of strong and weak coupling regimes. The flat space limit 
of our analysis of fluctuations about fuzzy sphere solutions should be related to 
the work in \cite{ohta1,ohta2}.

Another interesting avenue is to use the 
quadratic action we have obtained to do one and higher loop computations 
of the partition function and correlators. As indicated by the 
connections to integrability in section 2.2, these computations 
have interesting mathematical structure. It will 
also be interesting to incorporate the non-linearities in the fields 
perturbatively in the region of 
$R$ close to $R_0$.

\bigskip

{\bf Acknowledgments}:
We thank Antonis Antoniou,  Costas Kounnas,  Gabriele Travaglini and 
 John Ward for discussions.
 The work of SR is supported by 
a PPARC Advanced Fellowship. CP would like to acknowledge a
QMUL Research Studentship. This work was in
part supported by the EC Marie Curie Research Training Network
MRTN-CT-2004-512194.

\begin{appendix} 
\section{ Useful formulas for derivation of fluctuation
 action from D0-branes } 

A list of useful formulas is the following 
\bea\label{compcomsq}
 [\Phi_i,\Phi_j][\Phi_j,\Phi_i] &=& 8 N^2 \hat R^4 +8 \hat R^2
\left(\partial_a \phi\right) \left(\partial^a \phi\right) +48 \hat R^2
 \phi^2 +32 \hat R^3 N \phi - 48
\hat R^3 \frac{\epsilon^{ab}}{\sin{\theta}}F_{ab}\;\phi \nn\\
\nn && +32 \hat R^3 
\frac{\epsilon^{ab}}{\sin{\theta}}A_b\left(\partial_a \phi \right) -16 
\hat R^4 N 
\frac{\epsilon^{ab}}{\sin{\theta}} F_{ab}+16 \hat R^4 F_{ab}F^{ab}+64 \hat R^4  A_a  A^a\nn \\
 && -64 \hat R^4
\frac{1}{\sin^2 {\theta}}\left[ \epsilon^{ab} (\partial_{\theta}A_a) 
(\partial_{\phi}A_b )
  +A_{\phi}\left(\partial_{\phi}A_{\theta}\right) \cot{\theta} 
 \right] 
\eea
\be
 \left(\partial_t \Phi_i \right)\left(\partial_t \Phi_i \right) =
N^2 \dot{\hat R}^2 -4 \hat R^2 F_{0a}F^{0a}+\dot \phi^2 +2 \dot{\hat
  R}N \dot \phi + 4 (\dot{\hat R})^2 A_a A^a+4 \hat R \dot {\hat R}\;
\partial_t \left(A_a A^a \right)
\ee

\be
 ( \partial_t \Phi_i ) ~  [ \Phi_i, \Phi_j]  = 
2 i { \hat R } \dot{\hat R }N K_j^a (\partial_a \phi) + 4 i {\hat R }^3N 
    \epsilon_{ijp}x_p K_i^a (\partial_t A_a) 
\ee
\bea
\nn 
 (\partial_t \Phi_i) 
 [ \Phi_i, \Phi_j] [ \Phi_j , \Phi_k ]  (\partial_t \Phi_k) &=& 4 \hat
 {R }^2   \dot{\hat  R }^2 N^2\hat{h}^{ab}
( \partial_a \phi )
  (\partial_b \phi )  + 16 { \hat R }^6N^2\hat{h}^{ab} ( \partial_t A_a )
(\partial_t A_b )  \nn \\ && 
\nn  + 8 \dot{\hat  R } { \hat R }^4N^2 \omega^{ab}  ( \partial_t A_a )
 ( \partial_b \phi )     - 8 \dot{ \hat R}  { \hat R }^4N^2 \omega^{ab} 
( \partial_a \phi )  ( \partial_t A_b ).\\ 
\eea 

   To get the quadratic fluctuations we take a square root, 
expand, use the matrix correspondence between the trace and the
integral over the sphere (\ref{stdsph}), 
 and also employ the equations of motion. 
Note that, after taking the trace, 
 the terms in the last line in (\ref{compcomsq})
will combine with the linear term  $ -16 
\hat R^4 N \frac{\epsilon^{ab}}{\sin{\theta}} F_{ab} $ to give 
\be\label{td}  
-16 
\hat R^4 N  \epsilon^{ab}  \left ( 
F_{ab}  + i [ A_a , A_b ] + { 2 \over N } (\partial_{c}\omega^{cd} ) 
 ( A_a \partial_d A_b )  \right )  \;.
\ee
 We see that  $ F_{ab}$ gives a total derivative while 
the last two terms are not individually total derivatives 
but combine as such. The need for additional 
terms in the field strength, beyond  the commutator 
$[ A_a , A_b ]$  ( defined in (\ref{comm}) )  was explained in 
\cite{iktw}.  The  terms in (\ref{td})
 can be neglected when we are considering topologically trivial fluctuations.

\end{appendix}

\end{document}